\documentclass{aa} 
\usepackage[a4paper, left=1.5cm, right=1.5cm, top=2.5cm, bottom=2.5cm]{geometry}
\usepackage{amssymb}

\usepackage{epic,eepic}
\usepackage{graphicx}
\usepackage{color} 

\usepackage{natbib}
\bibpunct{(}{)}{;}{a}{}{,} 

\usepackage[english]{babel}

\usepackage[latin1]{inputenc}
\usepackage[T1, OT1]{fontenc}

\newcommand{\superscript}[1]{\ensuremath{^{{#1}}}}

\newcommand{\mgii}{Mg\,{\sc ii}}
\newcommand{\civ}{C\,{\sc iv}}
\newcommand{\ciii}{C\,{\sc iii]}}
\newcommand{\mgi}{Mg\,{\sc i}}
\newcommand{\feii}{Fe\,{\sc ii}}
\newcommand{\oiii}{[O\,{\sc iii}]}
\newcommand{\hi}{H\,{\sc I}}

\title{Constraints on the relative sizes of intervening Mg II-absorbing clouds and quasar emitting regions}
\author{Daniel Lawther\inst{1} \and Troels Paarup\inst{1} \and Morten Schmidt\inst{1} \and Marianne Vestergaard\inst{1,2} \and Jens Hjorth\inst{1} \and Daniele Malesani\inst{1}}

\institute{Dark Cosmology Centre, Niels Bohr Institute, University of Copenhagen, Juliane Maries Vej 30, DK-2100 Copenhagen, Denmark \and Steward Observatory, University of Arizona, 933 N. Cherry Avenue, 85721 Tucson, AZ, USA}

\begin{document}

\abstract{A significantly higher incidence of strong (rest equivalent width $W_{r}>1$\,\AA) intervening \mgii\,absorption is observed along gamma-ray burst (GRB) sight-lines relative to those of quasar sight-lines. A geometrical explanation for this discrepancy has been suggested: the ratio of the beam size of the source to the characteristic size of an \mgii\,absorption system can influence the observed \mgii\,equivalent width, if these two sizes are comparable.}{We investigate whether the differing beam sizes of the continuum source and broad-line region of Sloan Digital Sky Survey (SDSS) quasars produce a discrepancy between the incidence of strong \mgii\,absorbers illuminated by the quasar continuum region and those of absorbers illuminated by both continuum and broad-line region light.} {We perform a semi-automated search for strong \mgii\,absorbers in the SDSS Data Release 7 quasar sample. The resulting strong \mgii\,absorber catalog is available online\thanks{The \mgii\,absorber catalog, along with a 
table describing the survey redshift coverage $g(z)$, is available in electronic form from {{http://cdsweb.u-strasbg.fr/cgi-bin/qcat?J/A+A/}} or via {http://dark.nbi.ku.dk/research/archive/}}. We measure the sight-line number density of 
strong \mgii\,absorbers superimposed on and off the quasar \civ\,$\lambda$ 1550 and \ciii\,\,$\lambda$ 1909 emission lines. } {We see no difference in the sight-line number density of strong \mgii\,absorbers superimposed on quasar broad emission lines compared to those superimposed on continuum-dominated spectral regions. This suggests that the \mgii\,absorbing clouds typically observed as intervening absorbers in quasar spectra are larger than the beam sizes of both the continuum-emitting regions and broad line-emitting regions in the centers of quasars, corresponding to a lower limit of the order of $10^{17}$ cm for the characteristic size of an \mgii\,absorbing cloud.}{}

\titlerunning{Relative sizes of \mgii-absorbing clouds and quasar emitting regions}
\maketitle

\section{Introduction \label{sec:intro}}

Both quasar and GRB spectra sometimes display intervening absorption lines. The \mgii\,$\lambda\lambda$\,2796, 2803 doublet is identifiable even in low-resolution spectroscopy and is therefore well-suited for statistical studies. Intervening \mgii\,absorption systems are associated with galaxy haloes, and may be due to galactic outflows produced by supernovae, as they do not appear to be virialized \citep{Bouche2006}. 
Intervening absorption systems may therefore trace the star formation rate of their host galaxies. For example, the equivalent width of strong (rest-frame equivalent width, $W_r$, greater than $1$\,\AA{} for the 2796\,\AA\,component) \mgii\,absorbers appears to be correlated with the \oiii\,luminosity of their associated galaxies \citep{Menard2011}, making them possible tracers of star formation rate at high redshift \citep{Matejek2012}. Alternatively, intervening \mgii\,lines could represent a reservoir of cool ($T\approx 10^4$ K) gas in galaxy haloes, providing raw materials for galaxy growth \citep{White1991} -- in this case at least some of the absorption systems should trace inflowing material. \citet{Kacprzak2011} suggest a bimodal \mgii\,absorber population where weak absorbers trace infalling gas and strong absorbers trace outflows. 

\citet{Prochter2006} report a surprising discrepancy in the statistics of \mgii\,absorption along quasar and GRB sight lines, finding roughly four times as many strong intervening absorbers in GRB spectra. This is puzzling, as both quasar and GRB sight-lines are \textit{a priori} assumed to be randomly distributed. \citet{Vergani2009} study a longer GRB-spectrum redshift path ($\Delta z$=31.5) than previous work, yet they still find the sight-line number density of strong absorbers toward GRBs is a factor 2.1\,$\pm$\,0.6 higher than toward quasars. 
They are also able to extend their study to lower equivalent widths. Interestingly, they find that the number of weak absorbers  (0.3\,\AA\,$<$\,$W_r$\,$<$1\,\AA) 
is similar along quasar and GRB sight lines. \citet{Sudilovsky2007} find no statistical difference between the sight-line number density of \civ\,absorbers along quasar and GRB sight lines.

\citet{Prochter2006} discuss various explanations for their result: (a) dust obscuration by the galaxies associated with the absorption systems could alter the apparent magnitudes of those quasars richest in strong intervening absorbers, causing them to be excluded from magnitude-limited quasar samples; (b) a subset of absorption systems in the GRB sample could be intrinsic to the GRB event (implying very large outflow velocities if we are to observe them as intervening - see also \citealt{Cucchiara2009}, who find no significant difference between the properties of strong \mgii\,absorbers in GRB and quasar spectra); or (c) there could be a selection bias towards detecting those GRBs that are gravitationally lensed by intervening galaxies, leading to a higher instance of absorption systems towards the GRBs that we detect compared to random sight-lines (as the \mgii\,absorbers are associated with galaxies). \citet{Frank2007} suggest a model based on the relative sizes of \mgii\,absorbing clouds to quasar and 
GRB 
emitting regions to explain the \citet{Prochter2006} result -- see \S~\ref{sec:dilution}.

\subsection{Implications of dust extinction}

The galaxies associated with \mgii\,absorbers may both extinguish and redden background quasars. A few recent studies have addressed this issue.
\citet{Menard2008} investigate whether these effects could cause quasars with strong intervening absorbers to drop out of quasar catalogs due to selection effects -- for example, the SDSS DR7 catalog is limited to objects with absolute $i$-band magnitudes brighter than $-22$ \citep{Schneider2010}, and quasar candidates are initially selected based on color cuts. They estimate that about 2\% of quasars with strong \mgii\,absorbers  drop out of the SDSS quasar sample due to a combination of extinction and reddening associated with the \mgii\,absorber alone. This selection effect is strongly dependent on the equivalent width of the absorber -- they find that about 50\% of quasars with an intervening \mgii\,absorber with $W_r>6$\,\AA\,drop out of the SDSS quasar sample. Very strong absorbers are, however, rare (only 1.2\% of strong \mgii\,absorbers in the \mgii\,absorber sample described in \S~\ref{sec:properties} have $W_r>4$\,\AA - only 0.06\% have $W_r>6$\,\AA), so the effect of this bias on strong absorber 
statistics is negligible. 
The selection mechanism for GRBs is different: they are detected via the transient gamma-ray flux and accurately localized in x-ray and optical emission, allowing a follow-up spectrum to be obtained. In the case of `dark GRBs', a subgroup of GRBs with low optical flux relative to their X-ray flux \citep[e.g.,][]{Jakobsson2004}, it is not always possible to locate the burst accurately -- a high signal-to-noise follow-up spectrum is in any case difficult to obtain for these bursts. The optical faintness of dark GRBs is likely a consequence of dust in their host galaxies \citep[e.g.,][]{Kruhler2011}, and therefore unrelated to intervening \mgii\,absorption systems.

 \citet{Sudilovsky2009} perform Monte Carlo simulations investigating the effect of dust extinction by strong absorbers and find that only about 10\% of the quasar-GRB absorber discrepancy can be accounted for by this effect. 
\citet{Stocke97} find five strong \mgii\,absorbers in the spectra of ten BL Lac objects -- this is larger by a factor of 4-5 than the number of strong absorbers expected based upon quasar sight-lines. \citet{Berg2011} investigate strong intervening \mgii\,absorption for blazars in general (of which BL Lac objects are a subtype) -- they find a factor $2.2^{+0.8}_{-0.6}$ higher incidence of strong absorbers in blazar spectra compared to quasars. They argue that dust obscuration affects the quasar sample and not the blazar sample, as blazars are usually identified in bands that are not sensitive to dust obscuration. However, based on the dust obscuration quasar drop-out fraction of 2\% reported by \citet{Menard2008} they conclude that the resulting selection bias is too small to explain the discrepancy.

\citet{Budzynski2011} recently reexamined the dust obscuration bias in SDSS quasar samples, including an extra effect not discussed in the studies mentioned above: degradation of signal-to-noise ratio of heavily dust-attenuated quasar spectra. If \mgii\,absorbers are dusty, spectra with absorption lines will generally have a lower signal-to-noise ratio than those without, lowering the significance of detection of absorption lines. They find that a combination of signal-to-noise and dust obscuration could, in fact, account for a large part of the GRB-quasar absorber statistics discrepancy  (up to a factor of 2). 

\subsection{Implications of gravitational lensing}
\citet{Rapoport2011} examine the possibility that many of the GRBs we see are gravitationally lensed by foreground galaxies which also harbor 
\mgii\,absorbing clouds, giving rise to a bias in the GRB sample towards sight-lines with absorbers. Using archival Hubble Space Telescope \textit{(HST)} imaging, they report that galaxies are located systematically closer to the GRB sight lines than would be expected from a random distribution of sight lines. This is expected if the GRB sample has a large proportion of gravitationally lensed GRBs. \citet{Porciani2007} find that GRB afterglows with more than one \mgii\,absorber are, on average, a factor of 1.7 brighter in the optical band than the other afterglows. However, this result has a low statistical significance due to the lack of GRBs with more than one absorption system. They also suggest a multi-band magnification bias for GRBs: sources such as GRBs that are particularly bright in more than one band are more likely to be lensed, irrespective of the nature of the lens. If \mgii\,absorbers are more likely to lens GRBs than quasars due to this magnification bias, the result would be a subset of GRBs 
that would otherwise be too faint to be detected, all displaying \mgii\,absorption. \citet{Chen2009} find suggestive evidence for lensing of the GRB in one of the five systems displaying strong \mgii\,absorption in their sample.

\subsection{Implications of beam-size dilution\label{sec:dilution}}

\citet{Frank2007} propose a geometric model in which the different beam sizes of quasars and GRBs affect the fraction of \mgii\,systems observed as strong absorbers. 
The argument is as follows:
if a background source has a larger angular size than the \mgii\,absorbing cloud, some of the light will avoid the cloud completely, giving a `dilution' of the equivalent width compared to a beam where all photons pass through the absorbing cloud. 
They model this effect using spherically symmetric clouds of a fixed core radius $r_c$, with a power-law decline in column density outside this core radius, testing the total column density sampled by various beam sizes (relative to $r_c$). They show that, in the case of beams hitting the core centrally, small beams will sample a 
higher average column density than beams larger than the core. At an impact parameter $b=r_c$, this trend is weakened. At $b=5r_c$ the effect is reversed: while all beams sample a reduced column density compared to $b=0$, large beams sample a larger effective column density as they cover more of the core. For this beam dilution effect to explain the quasar-GRB discrepancy in absorber statistics, it requires that the beam size of quasars and the characteristic size of a \mgii\,absorbing cloud (or substructure) are comparable, while the beam sizes of GRBs must be somewhat smaller: if both GRB and quasar beam sizes are much smaller than the absorbing clouds, only a small fraction of \mgii-absorbed beams will sample the edge of the cloud, rendering the beam dilution effect insignificant. One prediction of this model is that absorbers in GRB spectra should vary in strength over time, as different parts of the GRB fireball are expected to dominate the emission at different stages of the burst. While such a 
variation has been reported for GRB 060206 \citep{Hao2007}, this claim was later refuted \citep{Thone2008,Aoki}. Variable-strength intervening absorbers do not appear to be a common phenomenon in GRB spectra, although the need for high-resolution, high signal-to-noise time sequenced spectra limits the amount of observations for which it is possible to detect such variations. \citet{Delia2010} examine the very bright GRB 080319B ($z = 0.937$), which displayed four intervening \mgii\,absorption systems in the redshift range $0.5 < z < 0.8$, and find no evidence for variability in the intervening lines.

Quasars are almost certainly powered by an accretion disk around a central supermassive black hole (e.g., \citealt{PetersonAGN}). The UV-optical continuum spectrum is expected to be emitted by the central part of the accretion disk. The broad emission lines are emitted by photoionized gas. The detailed structures, geometries, and sizes of the continuum-emitting and broad line-emitting regions (BLR) are not accurately known, as they are too compact to be resolved even for the nearest active galactic nuclei. However, the measured time delays in the response of the broad lines to continuum luminosity variations 
\citep[e.g.,][]{Peterson2008, Peterson2004} indicate that the BLR extends to significantly larger radii (of order light-weeks to light-months) than the UV-optical continuum (of order light-days). Much of the broad-line radiation 
is therefore emitted at large radii from the continuum source, giving a larger effective beam size for the quasar light (BLR plus continuum) at these wavelengths.

Depending on the relative redshifts of a given absorption system and the quasar, the \mgii\,line will either absorb the continuum emission alone or the continuum-plus-BLR emission. 
If the \citet{Frank2007} geometric model is the correct explanation for the different absorber statistics between GRBs and quasars, we also expect to see different absorber statistics for the different emission regions in the quasar system itself. The reasons are as follows. For the \citet{Frank2007} explanation to work, the `average' absorber in a quasar spectrum must suffer a dilution effect compared to a GRB-absorber. Therefore, the quasar continuum beam size must be large enough for the dilution effect to come into play. On the other hand, the continuum beam cannot be much larger than the absorbers, as we in that case would not expect to see any strong absorption in quasar spectra. This in turn implies that absorbers backlit by continuum plus BLR light would suffer an even larger dilution effect \citep{Pontzen2007}, as the BLR is larger than the continuum region.  A comparison of absorber statistics for those absorbers only receiving continuum light and those also illuminated by BLR light is therefore a 
test 
of the geometric explanation. 
Note that since the two beam sizes reside in the same quasar systems, selection biases in the quasar sample play no role.

\subsection{Outline of this study}

The aim of this study is to test for a
difference between the sight-line number density of strong \mgii\,absorbers backlit by BLR light and that of absorbers illuminated by continuum-dominated emission. We identify a sample of strong \mgii\,absorbers in SDSS DR7 quasar spectra \citep{Schneider2010}, measure the comoving sight-line number density of these absorbers (\S~\ref{sec:IDLscript}), and test it for a dependence on the quasar rest-frame wavelength of the absorbed light (\S~\ref{sec:properties}). The outcome of this study has implications for the characteristic sizes of \mgii-absorbing clouds relative to the sizes of quasar BLR and continuum-emitting regions. Detection of a significantly lower sight-line number density of strong \mgii\,absorbers projected onto the quasar broad \civ\,and \ciii\, emission lines would directly support the \citet{Frank2007} geometrical explanation for the quasar-GRB strong absorber statistics discrepancy found in \citet{Prochter2006}.

\section{Data: the Mg II absorber catalog }\label{sec:IDLscript}
\subsection{Method: identifying strong Mg II absorbers}

We searched the SDSS Data Release 7 quasar sample \citep{Schneider2010} for strong \mgii\,absorption. For this study we used the 1D spectra processed by the SpecBS algorithm (the Princeton reductions). These spectra are binned as $\log(\lambda)$ with a pixel width corresponding to 69 km s\superscript{-1}. The spectrograph has a wavelength-dependent resolving power of between 1850 and 2200.\\ 

{We developed a semi-automatic algorithm\footnote{The code is available from the authors on request.} to find strong \mgii\,absorbers in the spectra. This algorithm can be summarized as follows: an approximation $F_{s}(\lambda)$ of the quasar's intrinsic spectrum was first made via a combination of box-car median smoothing of the observed spectrum $F(\lambda)$ and Gaussian fits to residuals corresponding to the peaks of broad emission lines. We experimented with the size of the boxcar and found that a width of 121 pixels is a good compromise: the box must be large enough that the smoothed spectrum does not trace absorption features with strengths corresponding to strong \mgii\,absorbers, yet small enough that the median smoothing follows the broad wings of quasar emission lines adequately. This smooth spectrum serves as a reference spectrum against which candidate \mgii\,absorbers are selected. The standard deviation, $S(\lambda)$, of the residuals after subtraction of the smoothed spectrum (measured over a 
121-pixel box centered on 
each pixel) was then calculated. Any pixel with $F(\lambda) < (F_{s}(\lambda)-3.5 S(\lambda)$) and a signal-to-noise ratio (averaged over a 121-pixel box centered on the pixel in question) of 8 or above was considered for inclusion in an \mgii\,doublet. The latter constraint was imposed due to our line-finding algorithm becoming increasingly unreliable at lower signal-to-noise -- we set the cutoff at the lowest signal-to-noise ratio where the algorithm retrieved more than 99\% of strong absorbers in the \citet{Bouche2006} comparison sample described in section \ref{sec:completeness}, ignoring any absorption systems in the comparison sample located at spectral regions that fell below our signal-to-noise cutoff. Spurious identifications sometimes occurred with this signal-to-noise cutoff, however they were reliably flagged for visual inspection according to the criteria described below. Each of these candidate absorption features was fitted with a single Gaussian profile. Each absorption feature was then 
searched for additional structure within a 5-pixel ($345$ km s\superscript{\sc -1}, corresponding to 2.5 times the spectral resolution at 4000 \AA) interval, and refitted as a double-Gaussian composite if a non-negligible extra peak was found.\\

The resulting list of candidate absorption line positions was then searched for line pairs with a rest-frame wavelength separation of $\delta\lambda=7.178$\,\AA{} (in vacuum), corresponding to the separation of the \mgii\,$\lambda\lambda$ 2796,2803 doublet. These candidates were re-fitted simultaneously using a Gaussian for each line, with the wavelength separation between the two centroids fixed. The equivalent widths of the Gaussian line profiles were recorded.
We limited our search to wavelengths longer than $1250$\,\AA$(1+z_{\rm em})$ due in part to the severe contamination by Lyman-$\alpha$ absorbers shortward thereof and, in part, to the fact that our smoothed spectrum does not follow the quasar's Lyman-$\alpha$ emission profile adequately. 

Spectra with intrinsic broad absorption lines were discarded from the sample using the catalog produced by \citet{Shen2011} for two reasons. Firstly, it proved difficult to measure the equivalent width of a bona fide \mgii\,absorber superimposed on a BAL trough. Secondly, the automatic procedure occasionally misinterpreted broad \civ\,absorption as an \mgii\,doublet.

The expected doublet ratio for \mgii\,is $W_{2796}/W_{2803}\,\approx~2$ if the absorber is optically thin, and $W_{2796}/W_{2803} \approx 1$ in the optically thick limit. The initial constraint imposed on the ratio of equivalent widths for \mgii\,doublet candidates was $ 0.33<W_{2796}/W_{2803}<2.9 $. These limits were chosen to ensure that the automatic procedure includes all strong \mgii\,absorbers in the sample of \citet{Bouche2006}, even for cases where the initial measurement of $W_r$ for one of the doublet components was impaired by line blending. Doublets with $W_{2796}/W_{2803}< 1$ or $W_{2796}/W_{2803}>2$ were then flagged for visual inspection. Very strong absorbers ($W_{r}>4$\,\AA), absorbers with narrow Gaussian widths ($\sigma<0.3$\,\AA), and fits with large residuals within three pixels of the line center, were also flagged for visual inspection. Candidate systems at observed wavelengths between 7000\,\AA\,and 8000\,\AA\,were also inspected, as several SDSS spectra displayed artifacts of 
atmospheric-line 
removal in this interval which could be misinterpreted as \mgii\,absorption. 

In the relatively few cases where the \mgii\,doublet profile is not fully convincing to the eye (e.g. because one of the components is blended with another absorption line, making the line center uncertain), we searched for the presence of other metal absorption lines, such as \feii\,$\lambda$ 2600 and \mgi\,$\lambda$ 2852, at the same redshift.

The resulting \mgii\,absorber catalog comprises 10367 strong \mgii\,absorption systems, of which 2084 have been visually inspected.

\subsection{Completeness of the Mg II absorber catalog}\label{sec:completeness}
The completeness of the sample was tested against the SDSS-DR3 intervening \mgii\,absorber sample of \citet{Bouche2006} which comprises 1806 \mgii\,absorbers at redshifts $0.37<z_{\rm abs}<0.80$, of which 1269 have $W_{r}>1$\,\AA, in spectra of quasars with redshift $z_{\rm em}<3.193$. 
The original purpose of the \citet{Bouche2006} sample was to investigate the cross-correlation of the \mgii\,absorbers with the presence of galaxies. Their upper $z_{\rm abs}$ limit is due to a completeness requirement for their galaxy sample. Thus, the \citet{Bouche2006} comparison sample does not cover all of the $z_{\rm abs}$ range we search. However, the wide redshift range of their quasar sample ensures that their database contains representative examples of absorbers both blueward and redward of the \civ\,$\lambda$ 1550 emission line as well as absorbers superimposed on it. These wavelength regions are the focus of our investigation in \S\,\ref{sec:qsosup}. Our algorithm retrieved 99.2\% of the \citet{Bouche2006} absorbers with $W_{r}>1$\,\AA\,(as measured by our fitting procedure).
\citet{Quider2011} have released a catalog of \mgii\,absorbers in SDSS DR4. 
They search all targets flagged as quasars by the SDSS spectroscopic pipeline with an apparent $i$-magnitude brighter than 20, and find 17042 \mgii\,absorbers, of which 9942 have $W_r>1$\,\AA.
Our algorithm retrieves 96.2\% of the strong absorbers in that sample. Including the strong absorbers from the \citet{Quider2011} sample that our algorithm missed does not significantly affect the analysis of \S\,\ref{sec:qsosup}.\\

For this study it is crucial that we have a uniform sensitivity to strong \mgii\,absorbers superimposed on the quasar continuum  and on the broad emission lines, respectively. Emission lines generally have a higher signal-to-noise than the rest of the spectrum.
We tested the sensitivity by means of the detection limit $\sigma_{\rm det}(\lambda)$, defined as the integrated noise spectrum over a resolution element around each pixel divided by the flux integrated over the same resolution element 
\citep[e.g.,][]{Vestergaard03}. This gives a minimum equivalent width ($W_{r,\rm min} > 3 \sigma_{\rm det}$) that can be detected at that pixel.

The detection limit $\sigma_{\rm det}(\lambda)$ of absorbers  was measured for all pixels searched (i.e. all pixels fulfilling our S/N criterion) within the quasar rest-frame wavelength intervals 1450--1500\,\AA, 1522--1570\,\AA\,and 1650--1700\,\AA\,(i.e., bluewards of, superimposed on, and redwards of the \civ\,line, respectively), for the entire quasar sample. The cumulative distributions of $\sigma_{\rm det}$ for these intervals are shown in Figure~\ref{fig:sigma_det}.  Based thereon, we define the minimum rest-equivalent width $W_{r,\rm min}$ for each wavelength interval such that 95\% of the pixels searched in that interval allow the detection of an absorption line of strength $W_{r,\rm min}$ at a 3$\sigma$ significance.
The best case was for pixels superimposed on the \civ\,emission line, for which $W_{r,\rm min}=1.02$\,\AA. The worst case was for the interval 1450--1500\,\AA, for which $W_{r,\rm min}=1.16$\,\AA. Discarding all absorbers with $W_r < 1.16$\,\AA\,reduces the total sample size by 20\% without significantly affecting the analysis and conclusions discussed in  \S\,\ref{sec:qsosup}.\\

For \mgii\,absorption systems with one of the doublet components superimposed on the central peak of a quasar broad emission line there is an additional uncertainty in the equivalent width measurement, as the emission line profile fits were less reliable for partially absorbed lines. We found that absorption lines that were superimposed within 3\,\AA\,(in the quasar rest-frame) of the emission line peak adversely affected the Gaussian fit of the emission line profile. This typically led to an underestimation of the absorption line's equivalent width of order $0.1-0.2$\,\AA, with stronger absorption generally leading to a more severe underestimation. This underestimation caused a few marginally strong \mgii\,absorbers superimposed within 3\,\AA\, of the emission line peaks to drop out of our catalog.\\

To quantify this effect we interactively re-fit the emission line profile for all $0.8$ \AA\,$<W_r<1$ \AA\,absorption systems for which the \mgii\,2796\AA\,component was located within $3$\,\AA\,of a quasar's \civ\,emission peak. This led to the reexamination of 16 absorption systems. Ignoring the few strongly absorbed systems where the 'correct' emission line profile was difficult to discern by eye, our updated, improved measurements show that the automated algorithm on average underestimates the equivalent width of these absorbers by 0.07\,\AA. Four of the tested systems were revealed as $W_r>1$\,\AA\,absorbers after adjustment of the emission line fits. Based on Figure \ref{fig:sigma_det}, the completeness of our sample for absorbers near the quasar's \civ\,emission line with $W_r\approx 0.8$\,\AA\,is around 70\% -- this implies that around six strong absorbers on the \civ\,emission line peak are missed due to poor fitting of the emission line profile.\\ 

Another estimate of the amount of missed systems on the \civ\,emission line peak can be obtained by counting the systems within $3$ \AA\,of the \civ\,emission line peak that have $0.93$\,\AA$<W_r<1$\,\AA, i.e., all lines that, according to the equivalent width underestimation described above, would (on average) cross the $W_r>1$\,\AA\,threshold. There are 7 of these systems. Correcting for completeness (around 90\% at $W_r>0.93$\,\AA) gives an estimated 8 missed \mgii\,absorbers.\\

The effect of these missed systems on our result is discussed in section \ref{sec:qsosup}. The redshift path covered by this survey within 3\,\AA\,of the \civ\,emission line peak is only $0.3$\% of the total redshift path covered by the survey. Therefore the effect of this bias on the total path length density found for \mgii\,absorbers is negligible, even if a similar effect is present in previous surveys. In general we advise caution in the use of the equivalent widths quoted in our \mgii\,absorber catalog for absorbers that are close to emission line peaks and that are not flagged as having been visually inspected.

\begin{figure}
		\includegraphics[trim=1.5cm 0cm 0cm 0cm,clip=true,scale=0.37]{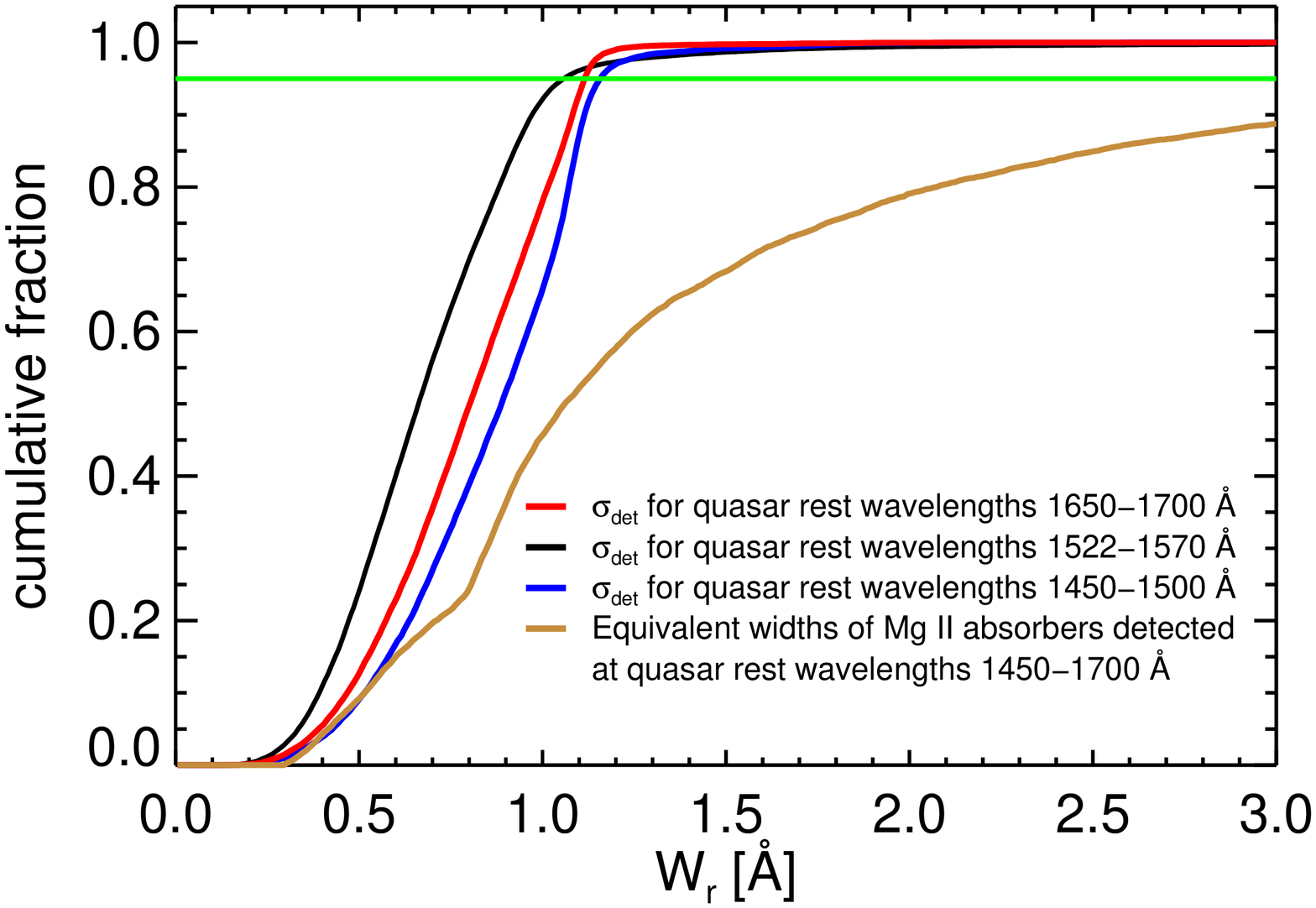}
	\caption{The cumulative distribution of the 3-sigma detection limit, $3\sigma_{\rm det}$, measured for all quasar spectra in our sample, for three quasar rest-frame wavelength intervals; the detection limit is defined in \S~\ref{sec:IDLscript}. Only the spectral regions in which we actually searched for absorbers (i.e. those that fulfill our signal-to-noise criterion, see \S~\ref{sec:IDLscript}) are included in the cumulative distributions. For comparison, the brown curve shows the cumulative distribution of equivalent widths of \mgii\,absorbers detected by our algorithm within the quasar rest-frame wavelength interval 1450-1700\,\AA. The horizontal green line shows the 95\% completeness limit.}
	\label{fig:sigma_det}
\end{figure}

\section{Results}\label{sec:properties}

In the following, we first establish the redshift path distribution of our \mgii\,absorber catalog (\S~\ref{sec:zdist}) and then examine the wavelength-dependent frequency of absorbers in the quasar frame of reference (\S~\ref{sec:qsosup}) in order to address the issue of the sizes of \mgii\,absorbers relative to the background quasar continuum and emission-line beam-sizes.  

\subsection{Redshift Distribution of Strong Mg II Absorbers}\label{sec:zdist}

The \mgii\,absorber catalog on which \citet{Prochter2006} base their comparison of GRB and quasar absorber statistics is that presented by \citet{ProchterQSO}, where 45023 SDSS DR3 quasar spectra were searched, resulting in 4835 strong \mgii\,absorbers. Before addressing the main question of the present study, it is therefore important to compare the properties of the \citet{ProchterQSO} absorber catalog to the properties of the absorber catalog presented here.  While our sample of quasar spectra is larger, we expect the statistical properties of the absorber populations to be similar: there are no obvious reasons to think that the \mgii\,absorber properties and typical quasar sight lines will be significantly different between the SDSS DR3 and DR7 releases. 

The redshift path density $g(z)$ of an absorber survey 
represents the number of spectra in which it is possible to detect strong \mgii\,absorption at a given redshift. A value of 1 is added to a $g(z)$ bin (corresponding to a redshift interval between $z$ and $z+\Delta z$ -- we set $\Delta z = 0.01$ in Fig. \ref{fig:gz} and $\Delta z = 0.15$ in Figures \ref{fig:obsXsnr8} and \ref{fig:appmags}) for each quasar spectrum that is searched for \mgii\,absorption at the corresponding redshift. Here, individual pixels in the spectrum were searched if the signal-to-noise ratio, averaged over a 121-pixel box centered on the pixel in question (corresponding to $\delta z=0.04$ when centered at $\lambda=4000$ \AA) was above 8. This often caused a fraction of the pixels in a redshift interval corresponding to a $g(z)$ bin to be skipped, while other pixels were searched for absorption -- in this case the fraction of the relevant redshift interval that was actually searched was added to the $g(z)$ bin. We show $g(z)$ as a function of redshift in Figure~\ref{fig:gz}. We 
attribute the dip in $g(z)$ around $z=1$ to a combination of a prominent sky emission line near 5570 \AA, for which the SDSS pipeline removal often results in a noise spike, and to the decrease in signal-to-noise of SDSS spectra near the wavelength corresponding to the split between the two CCD detectors. Either of these issues can cause the signal-to-noise in this spectral region to fall below our cutoff, thereby decreasing the effective redshift path length of our study. The sight-line number density $n(z)$ of strong \mgii\,absorbers in a given redshift bin is then $n(z)=N(z)/(g(z)\Delta z)$, where $N(z)$ is the number of absorbers found between redshifts $z$ and $z+\Delta z$, and $\Delta z$ is the size of a redshift bin.

\begin{figure}
		\includegraphics[trim=2.5cm 0cm 0cm 0cm,clip=true,scale=0.37]{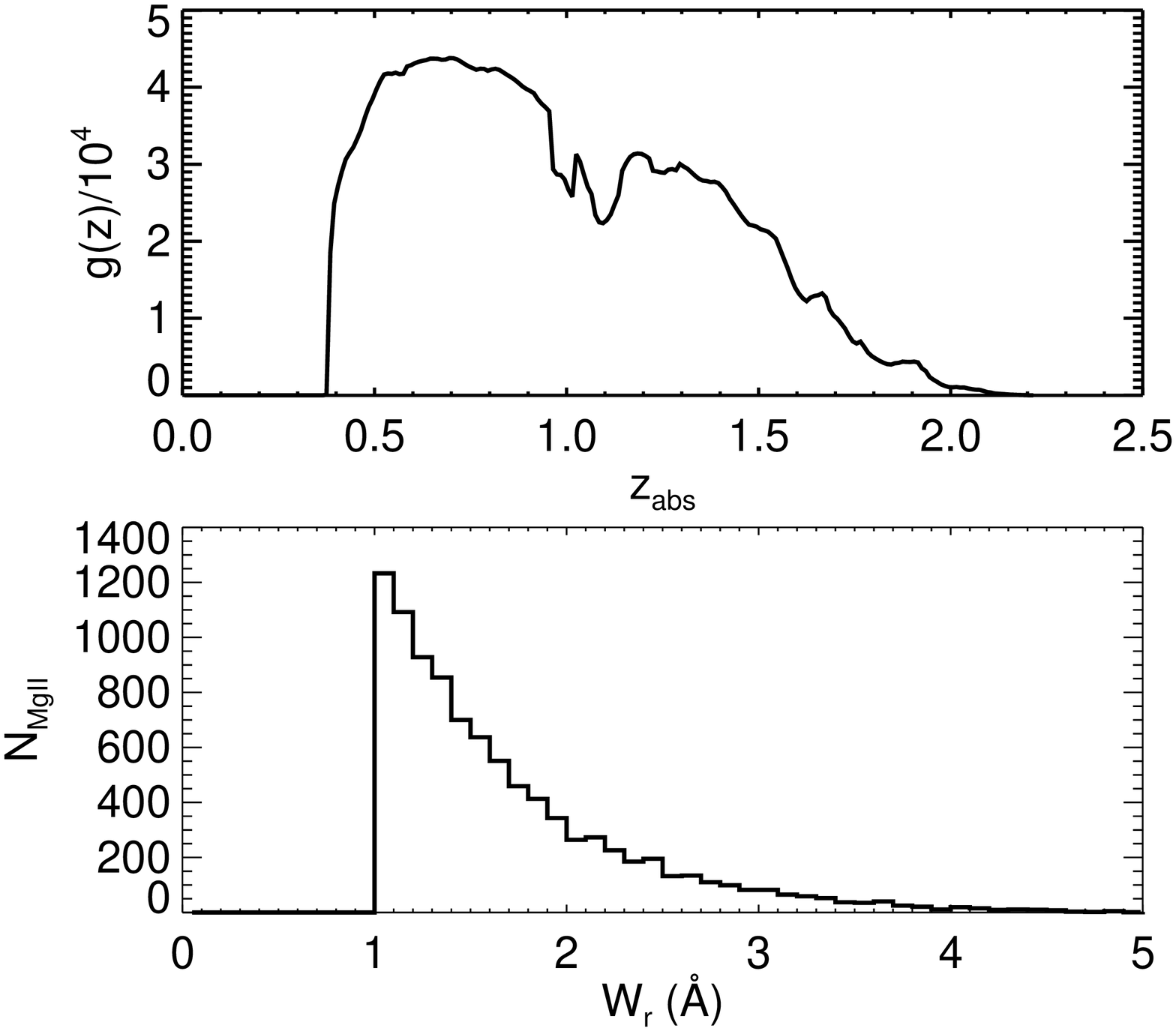}
	\caption{Top panel: the redshift path density $g(z)$ of this survey, i.e. the number of sight-lines that were searched for \mgii\,absorption for a given absorber redshift $z_{\rm abs}$. Bottom panel: the distribution of equivalent widths of the \mgii\,2796\,\AA\,component of the absorption systems in our sample.}
	\label{fig:gz}
\end{figure}

Following \citet{ProchterQSO}, we convert $n(z)$ to the sight-line number density of strong absorbers per comoving Mpc, $n(X)$: 

\begin{equation}
n(z)=n(X)(c/H_0)(1+z)^2[\Omega_M (1+z)^3+\Omega_{\Lambda}]^{-1/2}\rm,
\end{equation}
assuming a cosmology with density parameters $\Omega_{\Lambda}=0.73$, $\Omega_{M}=0.27$ \citep{Larson2011}. The redshift distribution of $n(X)c H_{0}^{-1}$\,is shown in Figure~\ref{fig:obsXsnr8}. 

\citet{ProchterQSO} find that an exponential relationship between redshift and $n(X)$ is a reasonable fit to their data, giving $n(X)cH_0^{-1}=n_0\exp(z_0/z)$ with $n_0=0.11(\pm 0.006)$, $z_0=-0.28(\pm0.05)$, for their $W_r>1$\,\AA\,sample. Fitting an exponential function to our data gives $n_0=0.11(\pm0.005)$, $z_0=-0.11(\pm0.03)$, i.e. we found a higher $n(X)$ than \citet{ProchterQSO} at low absorber redshift;  the exponential fits to $n(X)$ agree for both studies at $z\gtrsim1.0$ to within the quoted 1-$\sigma$ errors. When we integrate the fit to $n(X)$ of our absorber sample along the redshift path of GRB spectra searched for \mgii\,absorption in \citet{Vergani2009}, we obtain the number of \mgii\,absorbers that would be expected along quasar sight lines covering the same redshift path, $N_{\rm exp}=4.1^{+0.8}_{-0.7}$. This is consistent with the values of $N_{\rm exp}$ that \citet{Vergani2009} present in their table 3, based on the absorber samples of \citet{Prochter2006} and \citet{Nestor2005}, and 
approximately a 
factor 2 smaller than the sight-line density they find for GRBs.\footnote{Note that the $n(z)$ fit quoted in \citet{Prochter2006}, $n(z)=-0.026+0.374z-0.145z^2+0.026z^3$, is based on their updated absorber catalog covering SDSS Data Release 4. This fit is in better agreement with our measurement of $n(X)$ than that of \citet{ProchterQSO}, but the updated absorber catalog is not described in detail.}

\begin{figure}
		\includegraphics[scale=0.37,trim=1.5cm 0cm 0cm 0cm,clip=true]{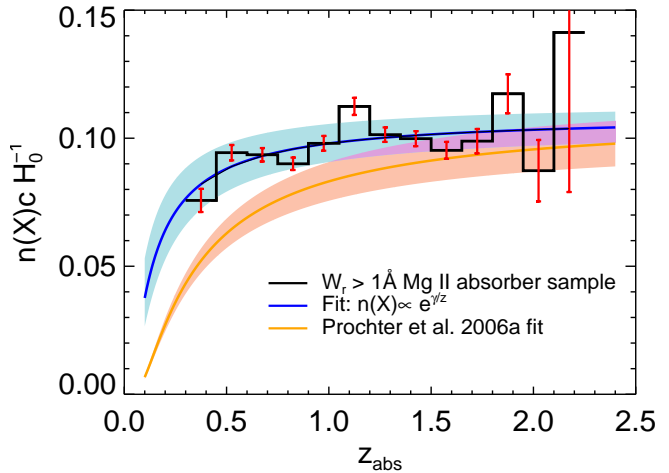}
	\caption{The redshift distribution of the sight line number density $n(X)$ of strong \mgii\,absorption systems per comoving Mpc (bins of $\Delta z=0.15$). The blue curve shows an exponential fit to this number density. The yellow curve shows the fit of \citet{ProchterQSO} to their strong \mgii\,absorber sample. The shaded areas show the $1\sigma$ uncertainties on the exponential fits.}
	\label{fig:obsXsnr8}
\end{figure}

We find that the sight-line number density of strong absorbers in our sample depends on the apparent magnitude of the background quasar. In the $u$ band (effective wavelength 3551\,\AA) a significantly higher sight-line number density of strong absorbers is found in spectra of faint ($m_u$$>$19.5 mag) quasars. For the $i$ band (effective wavelength 7481\,\AA)  the trend is reversed: we find a higher sight-line number density of absorbers in spectra of bright ($m_i$$<$18.2\,mag) quasars, see Fig. \ref{fig:appmags}. In the $g$ and $r$ bands the bright and faint quasar samples generally have the same sight-line number density of strong absorbers (within Poissonian errors). This appears to be analogous to the result presented by \citet{Menard2008} based on an investigation of extinction, reddening, and gravitational lensing effects for an SDSS DR4 \mgii\,absorber sample. They find apparent magnitude offsets for a sample of quasars with intervening \mgii\,absorption compared to an unabsorbed reference sample, 
with the amplitude and sign of the offset depending on the SDSS filter band-pass -- their results (described in \S~\ref{sec:SNRbias}) are consistent with our findings.

\begin{figure}
		\includegraphics[trim=1.2cm 0cm 0cm 0cm,scale=0.37]{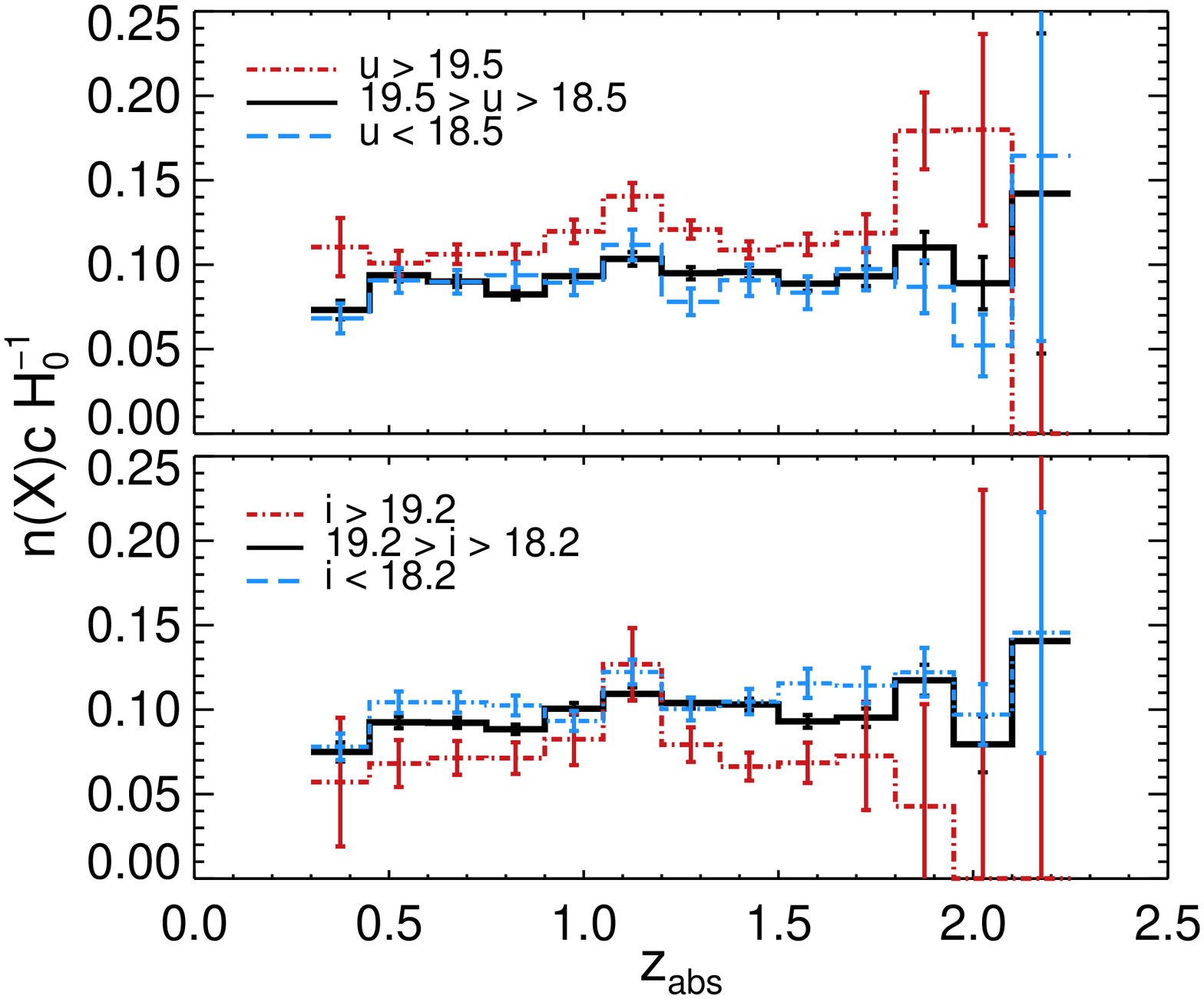}
	\caption{The redshift distribution of the sight line number density $n(X)$ of strong \mgii\,absorption systems per comoving Mpc (bins of $\Delta z$=0.15), divided into three subsamples according to the $u$-band (top panel) and $i$-band (bottom panel) apparent magnitude of the background quasar.}
	\label{fig:appmags}
\end{figure}

\subsection{Incidence of Strong Mg II Absorbers on Quasar Emission Regions}\label{sec:qsosup}

Here, we investigate the relative frequencies of \mgii\,absorbers superposed on the quasar continuum and broad line emission, respectively.
First, we define the superposition wavelength of each \mgii\,absorber as $\lambda _{\rm sup} = \lambda_{\rm obs} / (1+z_{\rm em})$, where $\lambda_{\rm obs}$ is the observed wavelength of the \mgii\,$\lambda$ 2796 component and $z_{\rm em}$ is the emission redshift of the background quasar. That is, $\lambda _{\rm sup}$ is the wavelength of the \mgii\,absorber in the quasar rest-frame. We split each quasar spectrum into bins of width 50\,\AA{} in the quasar rest-frame, and recorded the redshift path along which the \mgii\,absorption is searched in each bin for this sight-line. 
Summing this quantity over all spectra searched gives the redshift path density ($g\lambda_{\rm sup}$), analogously to $g(z)$ defined in \S~\ref{sec:zdist}, but now as a binned function of $\lambda _{\rm sup}$.

Next, we calculate the sight-line number density of strong \mgii\,absorbers for a given superimposition wavelength: $n(\lambda_{\rm sup})=N(\lambda_{\rm sup})/(g(\lambda_{\rm sup})\Delta z)$, where $N(\lambda_{\rm sup})$ is the number of absorbers found in each quasar rest-frame wavelength bin, $\Delta z$. We converted this quantity to the number density of strong absorbers per comoving Mpc, as in \S~\ref{sec:zdist} but now as a function of quasar rest-frame wavelength. We fitted a power-law $n\propto\lambda_{\rm sup} ^{\beta}$ to quasar rest-frame wavelength bins mostly free of BLR emission -- these bins are centered at 1320, 1355, 1452, 1690, 2030, 2160 and 2220\,\AA\,-- and found $\beta=0.13\pm0.11$. We also fitted a power-law to the entire $n(\lambda_{\rm sup})$ data set, yielding $\beta=0.17\pm0.10$. These fits are shown in Figure\,\ref{fig:qsorest}.

The \civ\,emission line in the composite SDSS quasar spectrum presented by \citet{VandenBerk2001} is centered at 1546.15\,\AA\ and has a full-width half-maximum (FWHM) of 23.78\,\AA. Here we regard \mgii\,absorbers with the quasar rest-frame wavelength of the 2796\,\AA\,doublet component located within this FWHM as being superimposed on the \civ\,line. The sight-line number density of strong \mgii\,absorbers superimposed on \civ\,is $n(X)cH_{0}^{-1}=0.101$. The power-law fit to the entire data set gave $n(X)cH_0^{-1}=0.089 $ at $\lambda_{\rm sup}$=1546\,\AA, while the power-law fit to the continuum-only quasar rest-frame wavelengths gave $n(X)cH_{0}^{-1}=0.088$ at $\lambda_{\rm sup}$=1546\,\AA. Adopting Poisson statistics, this corresponds to a significance of $1.08\sigma$ and $1.24\sigma$, respectively, for the peak in number density of absorbers superimposed on \civ\,compared to the whole-spectrum fit and continuum-only fit, respectively. 
In other words, we found no significant increase or decrease in the number of absorbers superimposed on the \civ\,emission line relative to that of the continuum emission. The slight overabundance of absorbers superimposed on the \civ\,peak could possibly be due to the higher signal-to-noise level at the position of the emission line, if the completeness of our search is significantly decreased at relatively low signal-to-noise levels near our cutoff of $\mathrm{S/N}>8$. This does not introduce a significant systematic error; see section \ref{sec:SNRbias}.\\ 

The sight-line number density of strong \mgii\,absorbers superimposed on the \ciii\, emission line (centered at 1905.97\,\AA\,with a FWHM of 21.19\,\AA) is $n(X)cH_{0}^{-1}=0.098$. The power-law fits gave $n(X)cH_{0}^{-1}=0.093$ and $n(X)cH_{0}^{-1}=0.090 
$ at $\lambda_{\rm sup}$=1906\,\AA\,for the whole-spectrum and continuum-only data, respectively. Adopting Poisson statistics, the significance of this peak is less than $1\sigma$ for both fits.\\

Due to the completeness considerations discussed in \S\,\ref{sec:completeness}, we repeat this analysis for absorbers with $W_r$$>$1.16\,\AA. In this case the significance of the overabundance of absorbers superimposed on \civ\,emission is reduced slightly, to $0.90\sigma$ and $1.06\sigma$ for the peak compared to the whole-spectrum and continuum-only fit, respectively. We also perform the same analysis including those absorbers from the sample of \citet{Quider2011} that our algorithm missed; doing so increased the significance of the \civ\,overabundance slightly, to $1.18\sigma$ and $1.27\sigma$ for the whole-spectrum and continuum-only fit, respectively.  Thus, these considerations do not change our main results summarized above.\\

As discussed in \S\,\ref{sec:completeness}, our algorithm misses some absorption lines superimposed on the central peak of the quasar broad emission lines, due to the absorption interfering with the emission line fitting. We estimate that around six strong \mgii\,absorbers superimposed on the \civ\,emission line are missed due to this issue. The addition of six extra absorbers at $\lambda_{\rm{sup}}=1550$ \AA\,increases the significance of the overabundance of absorbers superimposed on the \civ\,emission line to $1.48\sigma$.

\begin{figure}
	\centering
		\includegraphics[scale=0.37,trim=1.5cm 0cm 0cm 0cm]{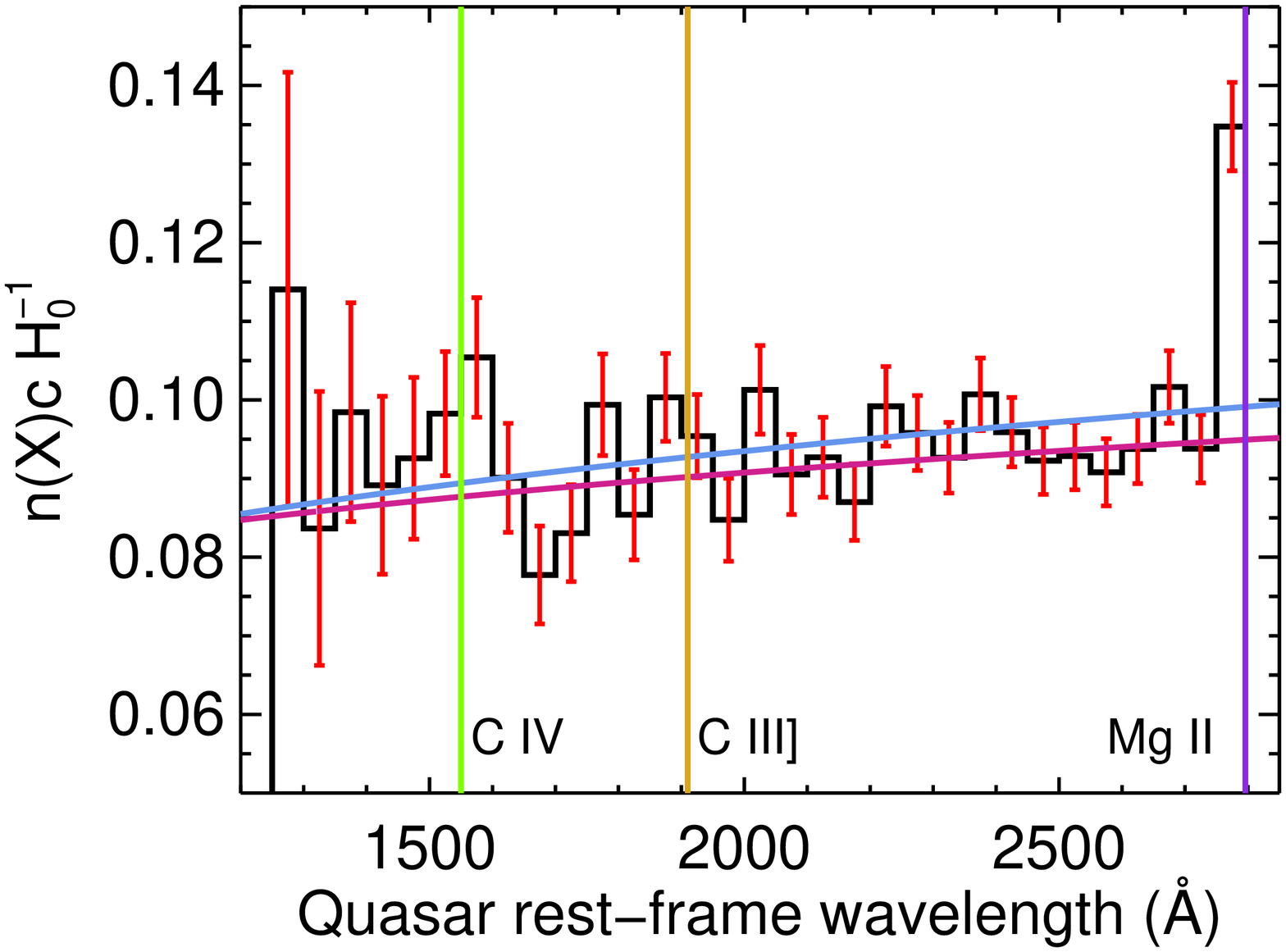}
	\caption{Distribution of the number density of strong \mgii\,absorbers per comoving Mpc with respect to the absorbers'  wavelength in the quasar rest-frame, $\lambda _{\rm sup} = \lambda_{\rm obs} / (1+z_{\rm em})$. We fit two power laws to the distribution: one is fitted to the number densities at continuum dominated wavelength bins only (violet curve), while the other is fitted to the entire dataset (blue curve). Error bars are Poissonian. Vertical lines indicate the wavelengths of the quasar broad emission lines \civ\,$\lambda$ 1550, \ciii\, $\lambda$ 1909, and \mgii\,$\lambda$ 2800. The peak at \mgii\,(i.e., $z_{\rm abs}\approx z_{\rm em}$) is expected, as some quasars display intrinsic \mgii\,absorption.}
	\label{fig:qsorest}
\end{figure}

\section{Discussion}\label{sec:discussion}

As outlined in \S\,\ref{sec:intro}, different source beam-sizes can alter the observed equivalent width of an intervening absorber. A simple model to illustrate this is to imagine all \mgii\,absorbing clouds as discrete, perfectly-absorbing spheres with solid boundaries. Consider a single \mgii-absorbing cloud on or near the quasar sight-line. In the limit of a point-source, we either observe a strong ($W_r>1$\,\AA) absorption or none at all. In the case of a background source (i.e., beam-size) much larger than the absorber radius, we will only observe weak absorption, as most photons from the source do not pass through the absorbing region at all. Between these two extremes we can observe strong absorption, depending on how close the absorbing cloud is to the quasar sight-line. For a random spatial distribution of sources and absorbers, a smaller source beam-size will therefore result in a larger number of strong absorbers than will a larger beam-size.

\subsection{Size estimates of source beams and Mg\,II absorbers}

In recent years, microlensing studies have been used to obtain approximate sizes of the continuum and line emitting regions in individual quasars. For example, \citet{Mediavilla2011} estimate the size of the continuum-emitting region of the quasar SBS 0909+532 ($z$\,=\,1.38), and obtain a radius of around $1.81\times10^{16}$ cm.  
For the quasar QSO 2237+0305 ($z$\,=\,1.7) \citet{Wayth2005} find a size of the \ciii\, emitting regions of about $1.85\times10^{17}$ cm and an upper limit on the continuum-emitting radius of $6.1\times10^{16}$ cm.
\citet{Mosquera2011} obtain microlensing measurements of 87 quasar continuum-emitting regions, finding radii of around $10^{15}$ cm. They also estimate the size of the H$\alpha$ BLR of the same quasars using the radius-luminosity relationship given by \citet{Bentz2009}, and find BLR radii of $\sim 10^{17}$ cm.
Owing to the ionization stratification of the BLR \citep{Clavel91}, higher ionization lines such as \civ\,are expected to be emitted closer to the ionizing continuum region than the \ciii\, and Balmer lines.
The size of the BLR scales with quasar luminosity \citep[see, e.g.,][]{Bentz2009}; the quoted figures serve as order-of-magnitude estimates of the relative sizes of continuum and \civ\,BLR in an `average' quasar in our sample. As an example of a quasar at $z\sim$\,2, \citet{Kaspi2007} report a tentative \civ\,broad-line rest-frame lag of 188$^{+27}_{-37}$ days with respect to variations in the continuum for the quasar S5 0836+71. This corresponds to a \civ\,BLR radius of $4.9\times10^{17}$ cm for this luminous quasar. \citet{Sluse2011} perform a microlensing study of the $z=1.695$ lensed quasar QSO 2237+0305 and find a radius of $1.71\times10^{17}$ cm for the \civ\,BLR, while \citet{Balashev2011} study the partial coverage of the quasar Q1232+082 ($z=2.57$) by an intervening damped Lyman-$\alpha$ absorber and find a \civ\,BLR radius of $2.78\times10^{17}$ cm, using photoionization considerations to estimate the size of the absorbing cloud. These considerations show that it is reasonable to assume 
that quasar BLRs are, in general, located at radii from the central black hole that are one or two orders of magnitude larger than the extent of the continuum-emitting region.\\

The radius of the GRB afterglow ring is expected to evolve with time \citep[e.g.,][]{Waxman97}. Based on this theoretical expectation \citet{Hao2007} estimate the radius of the GRB beam in GRB 060206 to be $5.1\times10^{15}$ cm at the time that their spectra were obtained. \citet{Frank2007} suggest that the effective beam size could be an order of magnitude smaller than this due to an uneven luminosity distribution over the GRB emission ring. This is smaller than the quoted size of the typical quasar continuum region by a factor of $\approx$10 or more.

While entire \mgii-absorbing complexes are thought to be similar in size to the gaseous haloes of their host galaxies \citep{Smette95}, there are signs of a clumpy substructure with individual clouds smaller than a kiloparsec in diameter as evidenced by lensed quasars \citep{Rauch02}.
This upper limit is very large compared to the beam-size estimates quoted above; thus the size of an \mgii-absorbing cloud itself is the main unknown in the present discussion.

\subsection{Our results in context of the beam-size dilution scenario}
In the context of the simple model described in \S~\ref{sec:dilution}, the beam-size interpretation \citep{Frank2007} implies that one of two relationships exists between the characteristic sizes of \mgii\,absorbing clouds and the background beams produced by (a) the quasar continuum emission, and by (b) the quasar BLR-plus-continuum emission, respectively.  One possibility is that the \mgii\,absorber size could be slightly smaller than the beam size emitted by a quasar continuum region. In this case, beam dilution always occurs for intrinsically strong absorbers in quasar spectra (regardless of the light source beam is the continuum or the BLR-plus-continuum), leading to the observed lack of strong absorbers along quasar sight-lines compared to GRB sight-lines.  Alternatively, the \mgii\,absorber could be comparable to or larger than the quasar continuum beam size but significantly smaller than the BLR beam size, in which case beam dilution only occurs for absorbers superimposed on BLR emission lines, and 
the discrepancy between quasar and GRB absorber statistics would be wholly due to a severe underabundance of strong absorbers superimposed on BLR emission.\\

In both of these  scenarios we expect to find an underabundance of strong \mgii\,absorbers superimposed on quasar \civ\,$\lambda 1550$ (and \ciii\, $\lambda$1909) emission lines compared to the quasar continuum, as the \civ\,(and \ciii\,) BLR extends to larger radii compared to the continuum region and contributes a significant amount of the total luminosity at 1550\,\AA. Figure~\ref{fig:qsorest} and \S~\ref{sec:qsosup} present a clear non-detection of such an underabundance. Therefore, the average size of \mgii\,absorbing clouds must be larger than the typical quasar continuum beam size. In fact the lower limit on the size of an absorbing cloud must be comparable to the BLR beam size, i.e., on the order of $10^{17}$ cm, as we see no evidence of beam dilution for absorbers superimposed on BLR emission. This result is in broad agreement with the analysis of \citet{Pontzen2007} for the \ciii\, emission line. A detailed comparison with the \citet{Pontzen2007} study is, however, difficult as they do 
not estimate the  completeness of their absorber sample (or, alternatively, test for a difference in their completeness on and off the \ciii\, emission line).\\

The simple picture described above is complicated somewhat by our ignorance of the true spatial distribution of the BLR gas, which is treated here as a spherical source of a given radius. Models in which the BLR emission is concentrated into smaller physical volumes (e.g. a rotating ring) for a given BLR radius may produce a less prominent dilution effect (averaged out over a large number of systems). Future quasar reverberation mapping studies aimed at retrieving the structure of the BLR \citep{Horne04} may give a better idea of the BLR beam morphology. However, we note that the BLR cannot be concentrated into an arbitrarily small spatial region with a negligible beam size, as the broad-line velocity distribution shows little structure at high resolution; this indicates that at least $5\times10^4$ broad-line emitting clouds orbit the central source \citep{PetersonAGN}.\\

\subsection{Density considerations for gas in a galaxy halo}

It is tempting to use the lower limit (of order $\approx 10^{17}$ cm) obtained in this study to place limits on the volume density of the clouds using column density estimates gleaned from the equivalent width of the \mgii\,absorbers, thereby constraining the conditions of the absorbing galaxy halo. Here we consider the implications of our constraint on \mgii\,absorption cloud size for the density in the clouds; to do so we must rely on knowledge of the cloud structure of \mgii\,absorbers inferred from high-resolution spectroscopy.\\

Due to the geometrical nature of the beam dilution effect, the lower limit on \mgii\,cloud sizes obtained in the present study ($\approx 10^{17}$ cm) in principle only applies to a single absorbing cloud. Most \mgii\,absorption systems in the SDSS display a saturated $2796,2803$\,\AA\AA\,doublet, and therefore the column density of strong \mgii\,absorbers in the SDSS cannot be determined from the absorption line profile alone \citep{Nestor2005}. However, \mgii\,absorbers with $W_r>1$ \AA\,as observed in moderate-resolution spectroscopy often turn out to be made up of a number of smaller, kinematically distinct components when seen in higher-resolution spectra. For example, \citet{Churchill2001} used Keck HIRES quasar spectra to study 23 \mgii\,absorption systems, of which five displayed strong ($W_r>1$ \AA) absorption. Their strong systems split into an average of 2.2 resolved components, with the strongest component of each system having an average $W_r$ of 1.09 \AA\,for the 2796\,\AA\,transition. 
Voigt profile fits suggest a further division of these resolved components. 
An extreme example is the strong \mgii\,absorption system at $z=0.911$ in the spectrum of quasar PKS 0823-223; this profile splits up into 18 Voigt profile cloudlets, with \mgii\,column densities $11.45<\log(N \mathrm{cm}^{-2})<13.25$ cm$^{-2}$.\\

{\citet{Menard2009} performed a study of the \hi\,content of \mgii\,absorbers. They generally found $N_{\textrm{\hi}}>10^{18}$ cm$^{-2}$ for \mgii\,absorbers with $W_r>0.45$\,\AA, with a large scatter -- see their Fig. 1. The median \hi\,column densities in their sample for \mgii\,absorbers with $W_r\approx 1$\,\AA\,is $3\times 10^{19}$ cm$^{-2}$. Assuming a single spherical absorption cloud and applying our lower limit for the cloud size, $\approx 10^{17}$ cm, this implies a \hi\,volume number density of $300$ cm$^{-3}$. The expected upper limit for the number density of gas in galaxy halos is $10$ cm$^{-3}$ \citep[e.g.,][]{Ding2005}. Our lower limit on cloud size is therefore a very weak constraint when considering a one-cloud model.\\ 

Alternatively one could assume that most strong \mgii\,absorption systems in the SDSS are comprised of several absorbing clouds, as the \citet{Churchill2001} results imply. The equivalent widths of individual cloud components could plausibly be of order $0.3$--$0.6$ \AA\,if the Voigt profile fits of \citet{Churchill2001} reliably reflect the substructure of strong \mgii\,absorbers. This corresponds to the smallest column densities found by \citet{Menard2009}, $N_{HI}\approx 10^{18}$ cm$^{-2}$. This column density gives a \hi\,number density of $10$ cm$^{-3}$ for our lower limit on \mgii\,cloud size, consistent with expectations for the densest halo gas \citep[e.g.,][]{Ding2005}.\\ 

\citet{Ding2005} performed photoionization modeling on weak and intermediate-strength \mgii\,absorption systems. For example, they estimate the size of the highest column-density velocity component in the $z=0.6600$, $W_r=0.34$\,\AA\,\mgii\,absorption system towards the quasar PG 1317+274 to be $6.17\times 10^{17}$ cm, with a \hi\,density of $\approx 2$ cm$^{-3}$; they estimate the size of the highest column density component of the $z=0.7729$, $W_r=0.694$ \mgii\,absorption system towards quasar PG 1248+401 to be $6.17\times 10^{18}$ cm, with a \hi\,density of $\approx 0.9$ cm$^{-3}$. Our lower absorber size limit is thus fully consistent with existing photoionization models for intermediate-strength \mgii\,absorption clouds, allowing for a scenario where strong absorbers in the SDSS are composed of multiple intermediate-strength components. It does not, however, strongly constrain such a scenario with respect to the sizes of the individual component clouds. According to the \citet{Ding2005} models, very 
weak \mgii\,absorbers will have kpc-scale sizes, much larger than the lower limit presented here.

\subsection{Bias due to low signal-to-noise?}\label{sec:SNRbias}

In \S~\ref{sec:zdist} we describe a dependence of the sight-line number density of strong absorbers on the apparent magnitude of the background quasar (Fig. \ref{fig:appmags}). \citet{Menard2008} find a similar dependence, and explain it as a combination of two effects: firstly a reddening of those quasar spectra with intervening absorbers due to dust associated with the absorption systems, and secondly a bias due to their line-finding algorithm. They measure this second effect using Monte Carlo simulations and find it to be independent of wavelength (i.e., producing a similar offset in all SDSS bands). 
This bias may be due to complications of continuum and emission line fitting in low-flux spectra -- such complications could affect the identification of \mgii\,doublets or the measurement of equivalent widths. But signal-to-noise may also be an issue as the ability of finding an absorber decreases in spectral regions with low flux; this presents itself as a tendency to find more (and weaker) absorbers in high-flux spectra (as discussed in \S~\ref{sec:properties}).  
Based on the discussion of completeness and sensitivity at the end of \S\,\ref{sec:properties}  this is not likely to affect our study given the high signal-to-noise cutoff ($\mathrm{S/N}>8$) and completeness level ($>$95\%) of our catalog. 
Nonetheless, we note that any potential bias of this nature in our algorithm must also be present to some degree in the \mgii\,absorber sample of \citet{Quider2011}, as the inclusion of absorbers from their catalog, missed by our algorithm, does not affect our main results.

\section{Summary\label{sec:summary}}

We have searched the SDSS DR7 quasar catalog for strong ($W_r > 1$\,\AA) \mgii\,absorption systems, and tested for a difference in sight-line number density of absorbers illuminated by quasar broad emission line photons compared to that of absorbers illuminated by continuum photons only. There is no significant difference in these number densities. This constrains the size of the \mgii\,absorbing clouds that give rise to strong absorption to be (as a lower limit) comparable to the size of quasar broad-line emitting regions, i.e., of the order of $10^{17}$ cm. The model proposed by \citet{Frank2007} to explain the difference in absorber sight-line number densities between quasar and GRB sight-lines appears to be excluded by our result as it implies an underabundance of absorbers on the \civ\,emission line relative to the continuum regions. This is in broad agreement with the test performed by \citet{Pontzen2007} for the \ciii\, broad emission line, and extends their result to the \civ\,broad emission line.

\acknowledgement
The Dark Cosmology Centre is funded by the Danish National Research Foundation. Funding for the SDSS and SDSS-II has been provided by the Alfred P. Sloan Foundation, the Participating Institutions, the National Science Foundation, the U.S. Department of Energy, the National Aeronautics and Space Administration, the Japanese Monbukagakusho, the Max Planck Society, and the Higher Education Funding Council for England. The SDSS Web Site is http://www.sdss.org/.\\


\bibliographystyle{aa} 
\bibliography{mgii} 
\end{document}